\documentclass[apjl]{emulateapj}

\newcommand{\xray}{\mbox{X-ray}}
\newcommand{\fig}{Fig.\,}
\newcommand{\tabl}{Table\,}
\newcommand{\eq}{eq.\,}
\newcommand{\mbi}[1]{\mbox{\boldmath$#1$}}

\shorttitle{Coma Cluster Galaxy Alignment} \shortauthors{Kitzbichler et al.}

\begin{document}

\title{Detection of Non-Random Galaxy Orientations in X-ray Subclusters of the Coma Cluster}
\author{M. G. Kitzbichler and W. Saurer}
\affil{Institute of Astrophysics, University of Innsbruck}
\email{manfred.kitzbichler@uibk.ac.at}
\email{walter.saurer@uibk.ac.at}
\begin{abstract}
This study on the Coma cluster suggests that there are deviations
from a completely random galaxy orientation on small scales. Since
we found a significant coincidence of hot-gas features identified
in the latest \xray\ observations of Coma with these local
anisotropies, they may indicate regions of recent mutual
interaction of member galaxies within subclusters which are
currently falling in on the main cluster.
\end{abstract}


%\keywords{galaxies: clusters: individual (A1656) --- galaxies:
%evolution --- galaxies: interactions --- galaxies: statistics ---
%X-rays: galaxies: clusters}

\section{INTRODUCTION}
Galaxy orientation studies in the past \citep[see][and references
therein]{1975AJ.....80..477H} were chiefly carried out in order to
verify miscellaneous galaxy cluster formation scenarios. Early
theories of cluster formation were the primordial-vorticity,
pancake-shock, and tidal-torque theories \citep[for a brief
overview see][]{1995MNRAS.276..327S} which had predicted different
alignment configurations of cluster members for different birth
models respectively. Unfortunately these models were at least
incomplete due to the missing knowledge of the presence of dark
matter in clusters and the assumption that clusters formed through
fragmentation. Putting the focus on finding an orientation pattern
for the whole cluster may be the main reason why many
investigations in this field did not yield sufficiently
unambiguous results
\citep{1986MNRAS.222..525F,1995MNRAS.276..327S,2001MNRAS.325...49F}.

More recent studies concentrated on the orientation of a number of
brighter galaxies whose major axis shows a tendency to be aligned
along the elongation of the parent cluster
\citep{1990AJ.....99..743S}. For instance the Virgo
\citep{2000ApJ...543L..27W} and also the Coma cluster
\citep{1997astro.ph..9289W} were found to exhibit this phenomenon
which is further discussed in section \ref{cdalign}.

Only recently \cite{2002MNRAS.333..501B} published their work that
tries to give an estimate on the reliability of gravitational
lensing studies which assume that anisotropies in the ellipticity
distribution are due to a gravitationally induced shear field
distorting galaxy images \citep[cf.][]{1990ApJ...349L...1T}, an
important aspect briefly picked up in section \ref{weaklensing} as
well.

In the present work, using extensive data on galaxy position
angles and inclinations of a large number of galaxies in Coma
(A1656), we are investigating the alignment of the majority of
faint cluster members on small scales, even though only in a
statistical way. Our method is to perform an individual
statistical test for the surroundings of each galaxy in the
sample. We show that the found anisotropies are statistically
significant to a high extent and superpose an image of average
orientations of non-random regions on recent \xray\ observations.
Since an interesting coincidence emerges this approach seems to be
profitable. We also try a tentative interpretation based on
current models of galaxy cluster evolution.

\section{THE DATA}
A huge sample of 6724 galaxies within a region of about
$2.6\degr\times 2.6\degr$ centered on Coma was compiled in
\citeyear{1983MNRAS.202..113G} by
\citeauthor*{1983MNRAS.202..113G} (GMP hereafter). This is still
the most comprehensive catalogue of galaxies in the Coma cluster,
comprising galaxies up to a brightness within the $26\fm 5$
contour limit of $b_{26.5}=21\fm 0$. The catalogue is considered
complete up to $b_{26.5}=20\fm 0$. Position angle and ellipticity
for 4344 entries are included, however no morphological
classification is given.

%\clearpage
\begin{figure}[htb]
\plotone{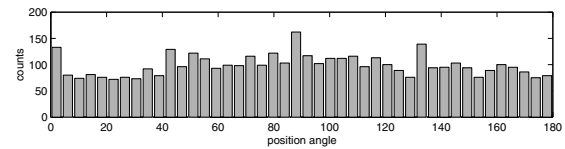}
\caption{Distribution histogram of the 4344
position angles in the GMP sample.\label{pahists}}
\end{figure}
%\clearpage

\label{90deghump}The sample as a whole shows a slight
overabundance of galaxies with position angles around $90\degr$
(see \fig\ref{pahists}) with a roughly harmonic shape of period
$180\degr$. Also a periodicity of $45\degr$ may be inferred which
was presumably caused by the plate scanning process but can be
neglected for our purposes since it is smoothed out by the binning
process. These global properties have yet been known for quite
some time \citep{1983ApJ...274L...7D}. However, a closer look at
the respective orientations of the individual galaxies is
worthwhile and provides new insights.

\section{THE METHOD OF ANALYSIS}
Instead of separating the cluster into certain sub-samples which
requires us to make assumptions on where we expect to be alignment
in the first place, we decided to use the environs of each galaxy
as a sub-sample which eliminates such a selection bias.

For each galaxy the angle of intersection between its major axis
and the major axes of the 15 closest neighbor galaxies is
calculated, whereby 15 was chosen as a reasonable trade-off
between small-number statistics and unwanted smoothing of the
results. From these differential angles a distribution histogram
is created. If the distribution is not in agreement with a random
distribution, i.e. it is not isotropic according to a
$\chi^2$-test, a best fit for the average orientation of this
ensemble of adjacent galaxies is derived as described below.

First, the \lq projected spin vector\rq~$\mbi{P}_i$ of a galaxy
with PA $\beta_i$ and ellipticities $\epsilon_i$ is defined as the
vector in the observational plane that is orthogonal to the major
axis of the galaxy image and has length $P_i=\sin(\eta)$. Here
$\eta$ denotes inclination, thus $cos(\eta)=\frac{b}{a}$,
$\epsilon=1-\frac{b}{a}$ with axial ratios $\frac{b}{a}$.
\begin{equation}
\mbi{P}_i = P_i\left(\cos\beta_i\atop\sin\beta_i\right)
\end{equation}
\begin{equation}
P_i^2=1-\frac{b^2}{a^2}=1-(1-\epsilon_i)^2
\end{equation}

For the ensemble we define the \lq average projected spin
vector\rq~$\mbi{S}$ with PA $\alpha$
\begin{equation}
\mbi{S}=S\left(\cos\alpha\atop\sin\alpha\right)
\end{equation}
as the vector which satisfies the condition
\begin{equation}\label{maxeq}
\left\{\sum_i(\mbi{S \cdot P}_i)^2\right\}\longrightarrow\max
\end{equation}
The scalar product can be expanded to
\begin{equation}\label{coseq}
(\mbi{S \cdot P}_i)^2=S^2P_i^2\cos^2(\alpha-\beta_i)
\end{equation}
Thus $P_i^2$ can be viewed as the weighting factor of the sum
which means that edge-on galaxies have a much stronger weight than
(nearly) face-on galaxies for which the PA is more or less
arbitrary.

It follows that \eq\ref{maxeq} has the solution:
\begin{equation}\label{alphatan}
\tan(2\alpha)=\frac{\sum\limits_iP_i^2\sin(2\beta_i)}{\sum\limits_iP_i^2\cos(2\beta_i)}=:\frac{S_y}{S_x}
\end{equation}
If we identify x and y in \eq\ref{alphatan} we get the length of
vector $\mbi{S}$ by computing $S^2=S_x^2+S_y^2$
\begin{equation}
S^2=\left(\sum_iP_i^2\cos(2\beta_i)\right)^2+\left(\sum_iP_i^2\sin(2\beta_i)\right)^2
\end{equation}

\section{RESULTS}
\begin{figure}[htb]
\plotone{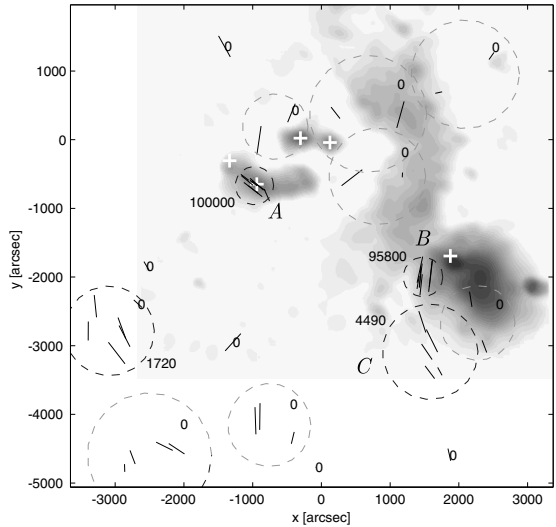}
\caption{Map of vectors $\mbi{S}$ for anisotropic
galaxy subsamples (\mbox{$\gamma_{\chi^2}$=0.99}). Dashed circles
denote clusters identified by K-means algorithm, numbers mean
quality $q$. The underlying image (extension marked by light grey
region) shows \xray\ residuals generated by subtracting a
$\beta$-model from the Coma cluster \xray\ image. White + symbols
denote massive galaxies. Origin at \mbox{$\alpha$=$12^h57\fm3$;}
\mbox{$\delta$=$+28\degr14\farcm4$} (1950.0), north is up, east is
left.\label{meanspinxoverlay}}
\end{figure}

The map seen in \fig\ref{meanspinxoverlay} shows the vector
$\mbi{S}$ plotted as black line at the position of each galaxy
which is found to possess an anisotropic neighborhood with a level
of confidence $\gamma_{\chi^2}=0.99$. It is interesting to note
that no significant anisotropies are found around the central cD
galaxies even though the galaxy density in this region is much
higher than anywhere else on the map. Moreover, clearly isolated
groups of galaxies with anisotropic orientation can be found.

Even though the algorithm favors such occurrence, the strength of
these features is greater than could be expected for a random
sample. In order to give an estimate of the statistical
significance of our results we performed a series of 1000 runs
with artificially generated isotropic data. Apart from being
impracticable due to the huge number of runs, a visual assessment
of these artificial plots would also be subjective therefore we
decided to apply a clustering algorithm to the distribution of
$\mbi{S}$. We chose a modified version of K-means clustering which
is simple to implement and fast enough for our purpose. The data
were clustered in a three-dimensional space spanned by the x and y
coordinates of $\mbi{S}$ and as a third coordinate by its PA
$\alpha$. The original algorithm had to be modified to be cyclic
in $\alpha$. Also we used as a distance measure:
\begin{equation}\label{clustdist}
r^2:=\Delta x^2+\Delta y^2+C\tan^2(\Delta\alpha)
\end{equation}
This definition makes sure that two $\mbi{S}$ vectors which have a
mutual PA difference of $90\degr$ are infinitely far apart as seen
by the clustering algorithm. The scaling factor $C$ can be viewed
as weight of $\alpha$. That is to say if $C$ is large the
clustering will be very sensitive to small differences in $\alpha$
and vice versa.

The artificial galaxy samples were generated by assigning random
PAs to each of the galaxies in the original sample. Moreover the
inclinations were shuffled among the sample members which has the
advantage of preserving the inclination distribution.

We performed two series of runs with 1000 artificial samples each.
The first one included the whole set of 4344 galaxies and a level
of confidence for the $\chi^2$-test of $\gamma_{\chi^2}=0.99$ was
chosen while for the second run only those 2354 galaxies brighter
than $b_{26.5}\le 20\fm 0$ were used which made it necessary to
decrease $\gamma_{\chi^2}$ to a value of 0.97 in order to get
enough $\mbi{S}$ vectors to work with (cf. Table~\ref{samptab}).

%\clearpage

\begin{deluxetable}{cccc}
\tablecaption{Definition of the samples\label{samptab}}
\tablecolumns{8} \tablewidth{0pc} \tablehead{\colhead{sample} &
\colhead{brightness limit} & \colhead{$N$}&
\colhead{$\gamma_{\chi^2}$}} \startdata
1 & $b_{26.5}\le 21\fm 0$ & 4344 & $0.99$ \\
2 & $b_{26.5}\le 20\fm 0$ & 2354 & $0.97$
\enddata
\end{deluxetable}

%\clearpage

The clusters produced by the K-means clustering had to be assessed
in an objective way thus we introduced a quality value $q$ for the
features found in the artificial samples as well as in the
original data:
\begin{enumerate}
\item Each feature that contains less than five members (i.e.
$\mbi{S}$ vectors) has $q=0$.

\item For each remaining cluster $q$ is defined as the norm of the
sum over its constituent $\mbi{S}$ vectors divided by its volume
in the three-dimensional cluster space.
\end{enumerate}
Table \ref{sigtab1} shows the results obtained for samples 1 and
2. It allows to compare the number $n$ and the quality $q$ of the
features in the data to the averaged values for these quantities
calculated from the set of artificial samples. Here $q_A$ denotes
the highest value for $q$ found in the data whereas $\bar q_A$ is
the average of the highest $q$ values calculated from each
individual artificial sample.
\begin{equation}
\bar q_A:=\sum(q_A^i)/1000
\end{equation}

%\clearpage

\begin{deluxetable}{ccccccccc}
\tablecaption{Properties of found anisotropies\label{sigtab1}}
\tablecolumns{9} \tablewidth{0pc}

\tablehead{\colhead{} & \multicolumn{ 2}{c}{data} & \colhead{} & \multicolumn{ 2}{c}{artificial} & \colhead{} & \multicolumn{ 2}{c}{significance}\\
\colhead{\raisebox{1.5ex}[-1.5ex]{sample}} & \colhead{$n$} &
\colhead{$q_A$} & \colhead{} & \colhead{$\bar n$} & \colhead{$\bar
q_A$} & \colhead{} & \colhead{$\sigma_n$} & \colhead{$\sigma_q$}}

\startdata
1 & 51 & $100000$ & & 29 & $16900$ & & 0.997 & 0.965\\
2 & 92 & $42900$  & & 71 & $10300$ & & 0.957 & 0.951
\enddata
\end{deluxetable}

%\clearpage

The rightmost two columns in \tabl\ref{sigtab1} give the
percentage of artificial samples that showed smaller values for
$n$ and $q_A$ than the data.
\begin{equation}\label{sigdef}
\sigma_x:=\frac{N(x_i<x)}{1000} \mbox{~~~with~} x \in\{n,q_A\}
\mbox{,~} i \in [1,1000]
\end{equation}

It becomes clear that only very few artificially generated samples
showed more or better clustered anisotropies than the original
data.

%\clearpage

\begin{deluxetable}{cccccccc}
\tablecaption{Qualities of top three clusters of a
sample\label{sigtab2}} \tablecolumns{10} \tablewidth{0pc}

\tablehead{\colhead{} & \multicolumn{ 3}{c}{data} & \multicolumn{ 3}{c}{artificial} & \colhead{}\\
\colhead{\raisebox{1.5ex}[-1.5ex]{sample}} & \colhead{$q_A$} &
\colhead{$q_B$} &      \colhead{$q_C$} & \colhead{$\bar q_A$} &
\colhead{$\bar q_B$} & \colhead{$\bar q_C$} &
\colhead{\raisebox{1.5ex}[-1.5ex]{$\sigma_q$}}}

\startdata
1 & $10^5$ & 95800 &  4490 & 16854 & 1054 &  19 & 1\\
2 &  42900 & 11000 & 10300 & 10340 & 2391 & 743 & 0.999
\enddata
\end{deluxetable}

%\clearpage

\begin{figure*}[htb]
\epsscale{.8} \plotone{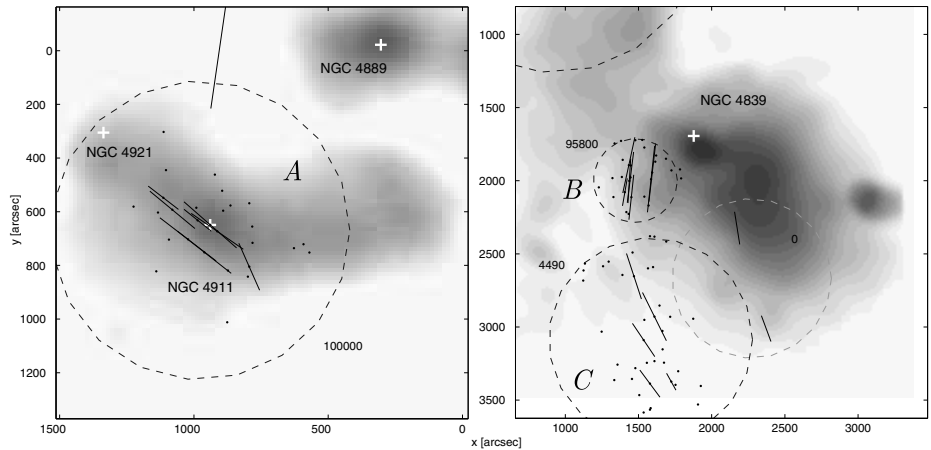}
\caption{Enlargements from
\fig\ref{meanspinxoverlay} showing the best three clusters $A, B,
C$ and their surroundings. Black dots denote those galaxies which
comprise the neighborhood of the $\mbi{S}$ vectors in the
cluster.\label{clusterdetail}}
\end{figure*}

In order to get a more sophisticated means of assessment we used
the three best clusters in a sample denoted $A$, $B$, and $C$ in
descending order of quality $q$. For the original data these would
be the very strong cluster around NGC4911 ($A$) and the slightly
less pronounced ones around NGC4839 ($B,C$). A detailed analysis
reveals that clusters $A$ and $B$ comprise 29 distinct galaxies
each whereas 39 distinct galaxies cause the anisotropic region
around cluster $C$. This can be seen in \fig\ref{clusterdetail}
which also shows the spatial distribution of those galaxies.

\tabl\ref{sigtab2} gives the results from comparing the cluster
qualities $q_A, q_B, q_C$ of the three top rated clusters in a
sample. Here
\begin{equation}\label{sigdef2}
\sigma_q:=\frac{N\left((q_A^i<q_A) \wedge (q_B^i<q_B) \wedge
(q_C^i<q_C)\right)}{1000}
\end{equation}
and $i$ is the index of the artificial sample as above. None of
the artificial samples could produce as many high quality clusters
of anisotropically distributed galaxies as the actual data.

%\clearpage
%\clearpage

In order to understand their physical meaning it is desirable to
verify the visually striking deviations from isotropy in
\fig\ref{meanspinxoverlay} by comparing them with other
measurements of anisotropy in the Coma cluster. An appropriate
means to this end can be the underlying image in
\fig\ref{meanspinxoverlay}. It shows the difference between the
original \xray\ intensity image of Coma and the expected intensity
for a so called $\beta$-model, which describes the distribution of
hot intra-cluster gas in a completely relaxed cluster. This
residual image was recently computed from a mosaic of XMM
observations of Coma composed by \citet{2003A&A...400..811N} who
kindly put their data at our disposal prior to publication.
According to them the pronounced feature in the SW around NGC4839
(white + symbol) is a subcluster currently falling on the main
cluster. Due to ram pressure inflicted by the intra-cluster gas
upon the subcluster, its gas is stripped off and leaves behind a
track of hot gas. The same mechanism seems to have produced the
slightly less strong residual around galaxies NGC4911 and NGC4921
located SE of the dominating cD galaxy pair in the center.

%\clearpage
%\clearpage

In \fig\ref{clusterdetail} indeed a striking coincidence is
observed between the regions of statistically significant galaxy
alignment and the features emerging in the residual image.
Moreover it may also be worth noting that the $\mbi{S}$ vectors
around NGC4839 seem to point in a direction tangential to its
perimeter while in the case of the NGC4911/NGC4921 pair the
anisotropies seem to coincide with the connecting line between the
two galaxies.

%\clearpage
\begin{figure}[htb]
\plotone{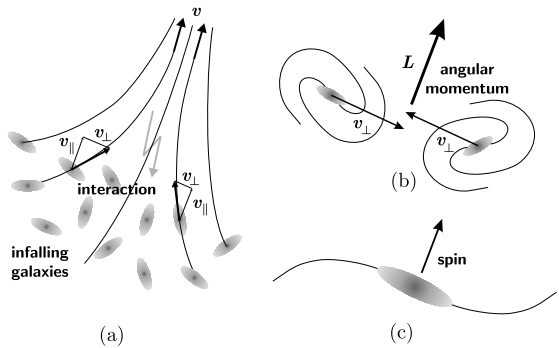}
\caption{The Laminar Flow Model: (a)~Stream of
infalling galaxies -- (b)~Orbiting galaxies -- (c)~Merger
product\label{LFM}}
\end{figure}
%\clearpage

\section{INTERPRETATION}
A possible explanation for the intriguing match found in
\fig\ref{meanspinxoverlay} and \fig\ref{clusterdetail} might be
given by a simple model of the dynamical evolution of interacting
galaxies falling on Coma along a filament, a process seemingly
very common for galaxy clusters
\citep{2002MNRAS.329L..47P,2002tceg.conf..395K,1999MNRAS.304L...5N}.
What we called the Laminar Flow Model (LFM) is outlined in
\fig\ref{LFM}.
\begin{enumerate}
\item Galaxies are streaming into the main cluster from outside.
The velocity vectors of two neighboring galaxies can be decomposed
into components parallel ($\mbi{v}_\parallel$) and perpendicular
($\mbi{v}_\perp$) to the velocity vector $\mbi{v}$ of the center
of mass (CoM).

\item If viewed in a co-moving reference frame with origin in the
CoM, the two galaxies orbit each other prior to merging. Their
angular momentum vector $\label{Ldef}\mbi{L}\sim\mbi{r \times
v}_\perp$ lies in a plane orthogonal to $\mbi{v}_\perp$ by
definition. Thus the projection of $\mbi{L}$ upon the
observational plane is statistically more likely to be aligned
parallel to the infall direction $\mbi{v}$.

\item Finally, after the merging has taken place, the resulting
galaxy tends to possess a spin vector parallel to the previous
angular momentum $\mbi{L}$.
\end{enumerate}
Basically the process can be viewed as funneling of galaxies along
the filaments. Perpendicular to the infall direction matter is
coalescing and getting denser whereas parallel to its motion
vector $\mbi{v}$ it is stretched due to tidal forces. Thus the
probability for a merger is higher if the component of the radius
vector $\mbi{r}$ parallel to $\mbi{v}$ is small. From the
definition of $\mbi{L}$ and \fig\ref{LFM}(b) it follows that the
observed spin alignment can be explained by the LFM. Moreover it
is in accordance with the view favored by
\citet{2002MNRAS.335..487M} which holds that the spin vector of
galaxies is chiefly determined by the last few merging events.

\label{cdalign}The ostensible contradiction with the findings of,
for instance, \cite{2000ApJ...543L..27W} that the major axes of
the brightest cluster galaxies are aligned with the filaments, can
be resolved by considering the different mechanism that produces
the bright central cD galaxies. They are the final destination of
the infalling matter whose linear momentum is thus converted into
angular momentum by the ultimate merger with one of those cosmic
cannibals.

\label{weaklensing}Finally the issue of weak lensing should not
remain unmentioned. On one hand a detected orientation anisotropy
in a sample of galaxies may be caused by gravitational lensing, on
the other hand if the anisotropies are real they may give wrong
lensing indications if not allowed for. These considerations are
beyond the scope of this paper but further investigation may be
worthwhile.

\section{CONCLUSION}
In this work we find that the overall isotropic appearance of
galaxy orientations in Coma can not be maintained when looking at
galaxy ensembles on the smaller scales of subclusters. These
fluctuations are not caused by statistical variations but there
exists strong evidence that they are the result of anisotropic
merging of subcluster members which fall on the main cluster
presumably along filaments extending between large clusters.
Therefore the identification of regions in which galaxies show
statistically significant deviations of their spin vectors from
isotropy may help to trace such filaments as well as provide a new
and useful tool to investigate the evolution of galaxy clusters.\\[2ex]
Acknowledgments: The mosaic of \xray\ observations of the Coma
cluster was provided by courtesy of D.~Neumann
\citep[cf.][]{2003A&A...400..811N}.
%\bibliographystyle{apj}
%\bibliography{../Diplomarbeit/references_new}

\end{document}